\documentclass[aps,twocolumn,pre,preprintnumbers,amsmath,amssymb,superscriptaddress,floats]{revtex4}

\usepackage{amsmath}	
\usepackage{amsthm}     
\usepackage{gensymb}
\usepackage{hyperref}
\hypersetup{colorlinks=true}
\usepackage{upgreek}
\usepackage{graphics}
\usepackage{hyperref}
\usepackage{epsfig}
\usepackage{color}
\usepackage{bm}
\usepackage{float}
\usepackage{ulem} 

\usepackage{url}
\usepackage{multirow}

\newcommand{\upperRomannumeral}[1]{\uppercase\expandafter{\romannumeral#1}}

\begin{document}

\title{Fidelity susceptibility of the anisotropic $XY$ model: The exact solution}
\author{Qiang Luo}
\email[]{qiangluo@ruc.edu.cn}
\affiliation{Department of Physics, Renmin University of China, Beijing 100872, China}
\author{Jize Zhao}
\affiliation{Center for Interdisciplinary Studies, Lanzhou University, Lanzhou 730000, China}
\author{Xiaoqun Wang}
\email[]{xiaoqunwang@sjtu.edu.cn}
\affiliation{Key Laboratory of Artificial Structures and Quantum Control (Ministry of Education), School of Physics and Astronomy, Tsung-Dao Lee Institute, Shanghai Jiao Tong University, Shanghai 200240, China}
\affiliation{Collaborative Innovation Center for Advanced Microstructures, Nanjing 210093, China}

\date{\today}

\begin{abstract}
We derive several closed-form expressions for the fidelity susceptibility~(FS)
of the anisotropic $XY$ model in the transverse field.
The basic idea lies in a partial fraction expansion of the expression
so that all the terms are related to a simple fraction or its derivative.
The critical points of the model are reiterated by the FS, demonstrating
its validity for characterizing the phase transitions.
Moreover, the critical exponents $\nu$ associated with the correlation length
in both critical regions are successfully extracted by the standard finite-size scaling analysis.
\end{abstract}


\maketitle
\section{Introduction} 
Quantum phase transitions~(QPTs), which are driven solely by quantum fluctuations and
are characterized by drastic changes in the ground state,
are of great interest for the interpretation of 
widespread phenomena in physics\cite{Sachdev2011,Vojta2003}.
Over the past decades, 
numerous theoretical methods for detecting QPT
have been introduced from the aspect of quantum information sciences
\cite{Osterloh2002,Zhu2006,Buonsante2007,ZhouZL2008,Garnerone2009l,Radgohar2018}.
The ground-state fidelity $F$\cite{Zanardi2006} is one of such methods. It measures the overlap between
two wave functions of the same Hamiltonian but at different values of the control parameter $\lambda$.
As a result, notable change in the fidelity is expected to occur at the transition point $\lambda_c$ even for a finite-size system.
However, the fidelity is sometimes chaotic numerically in that it depends on the increment
of the control parameter, and it vanishes exponentially with the increasing of the system size.
Therefore, the fidelity susceptibility~(FS) $\chi_F$\cite{You2007,Cozzini2007,CV2007}-the derivation of the fidelity
with respect to $\lambda$, is introduced to eliminate such drawbacks and turns out to be more powerful.
Technically, the FS is nothing but the nontrivial leading quadratic term of the fidelity,
so its divergence at the transition point is the reminiscence of the singularity of the latter.
The last 15 years has witnessed the explosive applications of fidelity and FS
to the QPT of various strongly correlated systems
\cite{Tzeng2008,ZhouB2008,Gu2010,Li2012,Lacki2014,Wang2015x,Wang2015l,Amiri2015,Luo2017,Sun2017,Wei2018},
including the intricate Berezinskii-Kosterlitz-Thouless transition
\cite{Yang2007,Wang2010,Sun2015}
and the unconventional topological phase transition
\cite{Abasto2008,Yang2008,Zhao2009,Garnerone2009a,Luo2014,Konig2016}.

Historically, the plausible evidence of the FS as a probe for quantum criticality
is revealed by the similarity between the scaling behavior of the FS at the critical point
and that of the second derivative of the ground-state energy, where the transverse-field
Ising model~(TFIM) is illustrated as an example\cite{Chen2008,Yu2009}.
Following standard arguments in the scaling theory of a continuous QPT, one obtains that the FS per site
in $d$-dimensional system with length $L$ scales as\cite{Gu2008,Albuquerque2010,You2011}
\begin{equation}\label{ChiScale}
{\chi_F(\lambda)}/{L^d} \sim L^{2/\nu-d}f\big(|\lambda-\lambda_c|L^{1/\nu}\big)
\end{equation}
where $\nu$ is the critical exponent of the correlation length and $f(\cdot)$ is a scaling function.
It is inferred that near the critical point the scaling expression
behaves as $\chi_F(\lambda_c)/L^d \sim L^{k_1}$~$(k_1=2/\nu-d)$.
Alternately, one may look at $\chi_F$ slightly away from the critical point at the thermodynamic limit~(TDL),
where $\chi_F(\lambda)/L^d \sim |\lambda-\lambda_c|^{k_2}$~$(k_2=d\nu-2)$.
Consequently, the scaling ansatz in the system exhibiting logarithmic divergences requires that
the absolute value of the ratio $k_2/k_1$ is the critical exponent $\nu$\cite{Gu2008,Albuquerque2010,You2011}.
The critical exponent $\nu$ is usually calculated numerically
because of the absence of analytical expression for FS.
The main obstacle lies in that no algebraic technique is available
to obtain its closed-form at finite length.
Breakthrough was made on the single-parameter TFIM by Damski \textit{et~al.} in light of
several elegant summation formulas\cite{Damski2013,Damski2014,Damski2015}.
This method, however, is difficult to follow and can hardly be generalized to
a double-parameter $XY$ model where a quadratic summation is explicitly involved\cite{Zanardi2006}.
Notably, after a proper symmetry analysis, we find that the partial fraction expansion
method can be used to divide the quadratic term into two coupled linear terms.
As a result, all the terms are related to a simple fraction or its derivative
~(see Appendix~\ref{AppA} and ~\ref{AppB})
that can be treated exactly in principle.

The remainder of the paper is organized as follows. In Sec.~\ref{SecModel} we introduce the
anisotropic $XY$ model in the transverse field, and give the expressions for the FS.
The closed-form of them are presented in Sec.~\ref{SecExact} in detail,
and the critical exponent $\nu$'s are calculated analytically by the scaling ansatz.
Sec.~\ref{SecConc} is devoted to the conclusion.

\section{Model}\label{SecModel}
We are going to calculate analytically the FS of the one-dimensional spin-1/2 anisotropic $XY$ model in the transverse field\cite{Lieb1961},
\begin{equation}\label{XYModel}
\hat{H} = -\sum_{n=1}^{\mathcal{N}}\Big(\frac{1+\gamma}{2}\sigma_n^x\sigma_{n+1}^x + \frac{1-\gamma}{2}\sigma_n^y\sigma_{n+1}^y + h\sigma_n^z\Big),
\end{equation}
where $\sigma_n^{\alpha}$~$(\alpha = x,y,z)$ is
the $\alpha$ component of the Pauli operator acting on site $n$,
$\gamma$ is the anisotropy parameter at the $xy$ plane,
and $h$ is the external field along the $z$ direction.
The number of sites $\mathcal{N}=2N$ is assumed to be \textit{even} for brevity
and both symbols $N$ and $\mathcal{N}$ are used throughout the paper.
Here periodic boundary condition~($\sigma_{\mathcal{N}+1}=\sigma_1$) is imposed
so that Eq.~\eqref{XYModel} can be diagonalized exactly via the Jordan-Wigner transformation.
The physical ground state depends highly on the proper choice of momenta quantization,
which results in a positive or negative parity sector\cite{Pasquale2009,Okuyama2015}.
The ground state could be determined by a competition between the vacuum states of
the two parity sectors at given parameters\cite{Pasquale2009}.
Interestingly, The energy gap, obtained from the two states, shows a rather anomalous behavior\cite{Okuyama2015}.
For even $\mathcal{N}$, it is argued that the physical ground state lies in the positive sector
at least outside the disordered circle $h^2+\gamma^2=1$.
In the circle, the energy gap vanishes rapidly near $\gamma=0$.
So for simplicity we will only consider the momenta quantization in the positive sector hereafter.
The single particle energies are given by
$\Lambda_k = \sqrt{\mathbf{\epsilon}_k^2+\gamma^2\sin^2k}$,
where
$\epsilon_k=\cos k-h$, $k$=$(2n-1)\pi/\mathcal{N}$~($n$=1,2,$\cdots$,$N$)\cite{Pasquale2009,Okuyama2015}.
With $\cos\theta_k=\epsilon_k/\Lambda_k$ in mind, we have the analytic expressions for the FS as
$\chi^{(q)}(h,\gamma) = \frac{1}{4}\sum_{n=1}^{N}\left({\partial \theta_k}/{\partial q}\right)^2$
with $q = h$ and $\gamma$\cite{Zanardi2006}.
Consequently, The explicit expressions for the FS with respect to the two kind of QPTs~(see below) are\cite{Zanardi2006}
\begin{align}\label{DefXYFSh}
\chi^{(h)}(h,\gamma) &= \frac14\sum_{k>0}\frac{\gamma^2\sin^2k}{\Big[(\cos k-h)^2+\gamma^2\sin^2k\Big]^2}
\end{align}
and
\begin{align}\label{DefXYFSr}
\chi^{(\gamma)}(h,\gamma) &= \frac14\sum_{k>0}\frac{\sin^2k(\cos k-h)^2}{\Big[(\cos k-h)^2+\gamma^2\sin^2k\Big]^2}.
\end{align}
Actually, the FS defined above is none other than the diagonal element
of a more general concept, the quantum metric tensor\cite{Zanardi2007}.
The expression for the remaining off-diagonal element of it
is shown in the appendix~\ref{AppC}.
\begin{figure}[!ht]
\centering
\includegraphics[width=7.5cm, clip]{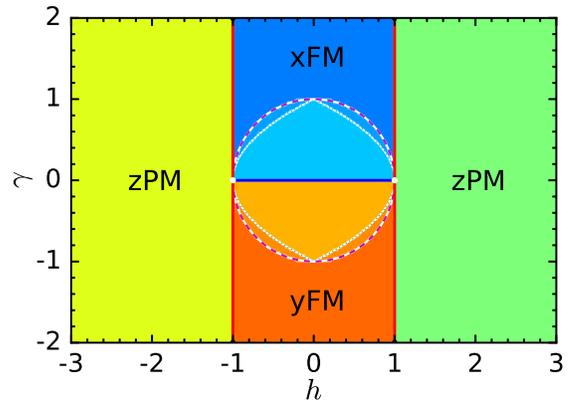}\\
\caption{(Color online) The ground-state phase diagram of the anisotropic $XY$ model in the transverse field at the $(h,\gamma)$ plane.
    There are paramagnetic phases with opposing orientation along the $z$ direction when $|h|>1$,
    and are ferromagnetic phases when $|h|<1$.
    There are two fascinating curves in the ferromagnetic phases~(see the main text)\cite{Lieb1961}.}\label{FIG-XYPD}
\end{figure}

As shown in FIG.~\ref{FIG-XYPD}, the model Eq.~\eqref{XYModel} has a richer phase diagram when compared with the TFIM,
a special case of Eq.~\eqref{XYModel} at $\gamma=1$.
There are four different phases in the $(h,\gamma)$ plane, which are
separated by the lines $h=\pm1$ and by the segment $|h|<1$, $\gamma=0$.  
The corresponding QPTs are referred as
the \textit{Ising transition} and the \textit{anisotropy transition}, respectively.
The Ising critical lines are the boundaries between ferromagnetic phases and paramagnetic phases,
whereas the anisotropy transition separates the ferromagnets with spins along the $x$ and $y$ direction.
Both Ising and anisotropy transitions share the same critical exponent $\nu=1$\cite{Cheng2010,Rams2011},
and we intend to calculate it analytically according to the expressions
Eq.~\eqref{DefXYFSh} and Eq.~\eqref{DefXYFSr} in the following section.
There are two fascinating curves in FIG.~\ref{FIG-XYPD}.
The circular curve~($h^2+\gamma^2=1$, pink color) separates the regions
with oscillatory and non-oscillatory correlations asymptotic behavior,
while the parabolic curve~($\gamma^2 \pm h=1$, cyan color) is the boundary
between commensurate and incommensurate phases\cite{Bunder1999}.
It is worth to mention that the critical exponents at the multicritical points
obey the different universalities\cite{Mukherjee2011} and is beyond the scope of current paper.

\section{Exact Solutions}\label{SecExact}
\subsection{The Ising transition}
Followed from Eq.~\eqref{DefXYFSh}, we see that the FS for $XY$ model
with respect to the external magnetic field $h$ can be rewritten as
\begin{align}\label{DefXYFShv1}
\chi^{(h)}(h,\gamma)
&= \frac{\gamma^2}{4(1-\gamma^2)^2}\sum_{n=1}^{N}\frac{1-c_n^2}{\Big[c_n^2- \frac{2h}{1-\gamma^2}c_n + \frac{h^2+\gamma^2}{1-\gamma^2}\Big]^2}
\end{align}
where $c_n = \cos\frac{(2n-1)\pi}{\mathcal{N}}$.
Here after $|\gamma|$$<$1 is assumed to avoid possible ambiguity,
and we note that our main results remain unchanged for arbitrary value of $\gamma$.
For example, our results are still valid for $|\gamma|$=1 since the FS is continuous when crossing the lines.
The summation of Eq.~\eqref{DefXYFShv1} is not easy to handle directly
due to the existence of the quadratic terms.
To eliminate them, we here employ a factorization method.
It can be found that when $h^2+\gamma^2>1$ we can factorize
$\phi(t)$=$t^2-\frac{2h}{1-\gamma^2}t + \frac{h^2+\gamma^2}{1-\gamma^2}$=$(t-\lambda_{+})(t-\lambda_{-})$
where $\lambda_{\upsilon}$'s~($\upsilon=\pm$) are the \textit{real} roots of the equation $\phi(t)=0$ and
\begin{align}\label{lambdaNu}
\lambda_{\upsilon} = \frac{h + \upsilon\vert\gamma\vert\sqrt{h^2+\gamma^2-1}}{1-\gamma^2}.
\end{align}
The sign of the two roots $\lambda_{\upsilon}$ are the same and is consistent with the sign of field $h$.
Besides, the absolute values of the roots are both larger than 1 so long as $|h|\neq1$
(in this case the smaller one equals to 1).
when $h^2+\gamma^2<1$, however, the roots are complex and the imaginary unit $i=\sqrt{-1}$ should be involved.
By virtue of the partial fraction expansion of the expression $\frac{1-c_n^2}{(c_n-\lambda_{+})^2(c_n-\lambda_{-})^2}$,
Eq.~\eqref{DefXYFShv1} is recast into the following symmetric form
\begin{equation}\label{FS-XY}
\chi^{(h)}(h,\gamma) = \frac{1}{16(h^2+\gamma^2-1)}\sum_{\upsilon=\pm}\mathcal{F}_{\upsilon}^{(h)}
\end{equation}
where
\begin{align}\label{FsumFldH}
\mathcal{F}_{\upsilon}^{(h)}
&= (\lambda_{\upsilon}^2-1)L'(\lambda_{\upsilon})-\frac{2(\lambda_{\upsilon}\lambda_{\bar\upsilon}-1)}{(\lambda_{\upsilon}-\lambda_{\bar\upsilon})}L(\lambda_{\upsilon})
\end{align}
with $\bar\upsilon$ be the complementary component of branch $\upsilon$~($\bar\upsilon\upsilon=-1$).
In Eq.~\eqref{FsumFldH}, the complex analytic function
$L(\alpha) = \sum_{n=1}^{N}\frac{1}{c_n+\alpha}$~($\alpha\in\mathbb{C}$) is introduced,
and the odevity of it and its derivative are utilized as well.
The highlight of the paper is that the closed form of $L(\alpha)$ is discovered,
making it possible to have exact expressions for FS.
The mathematical details are omitted here and are presented in the Appendix~\ref{AppA},
and the relation between the $L'(\alpha)$ and $L(\alpha)$ is shown in the Appendix~\ref{AppB}.
According to the formula presented in Appendix~\ref{AppA}, it is useful to have
$g_{\upsilon} = \lambda_{\upsilon}+\sqrt{\lambda_{\upsilon}^2-1}$, so that
\begin{align}\label{Def:gbeta}
g_{\upsilon} = \frac{h+\upsilon p_{\upsilon}\sqrt{h^2+\gamma^2-1}}{1-p_{\upsilon}|\gamma|}
\end{align}
where $p_{\upsilon}$=$p_{\upsilon}(h)$ is a piecewise sign function of field $h$ and branch $\upsilon$
\begin{align}\label{Def:qsgn}
p_{\upsilon} =
\left\{
  \begin{array}{ll}
    \upsilon, & |h|>1 \\
    \mathrm{sgn}(h), & |h|<1
  \end{array}
\right..
\end{align}
The relevant correlation length near the critical point reads $\xi=1/|\ln g_{+}|$\cite{Lieb1961},
indicating that it is the positive branch of $\mathcal{F}_{\upsilon}^{(h)}$~($\upsilon=1$)
that gives rise to the divergence behavior of the FS in the TDL.
The explicit expression of $\mathcal{F}_{\upsilon}^{(h)}$ is calculated analytically as follow:
\begin{equation}\label{Fsum}
\mathcal{F}_{\upsilon}^{(h)} = \frac{\mathcal{N}^2g_{\upsilon}^{\mathcal{N}}}{(g_{\upsilon}^{\mathcal{N}}+1)^2} + \frac{\mathcal{N}C_{\upsilon}}{2}\frac{g_{\upsilon}^{\mathcal{N}}-1}{g_{\upsilon}^{\mathcal{N}}+1}
\end{equation}
with the expression
\begin{align}\label{Def:Coeff}
C_{\upsilon} =\frac{p_{\upsilon}}{h^2-1}\left[\frac{\upsilon{\gamma}^2h}{\sqrt{h^2+\gamma^2-1}}-\frac{h^2+\gamma^2-1}{|\gamma|}\right].
\end{align}
We note here the detailed derivations are presented in the supplemental material~(SM)\cite{MySUPP}.

\begin{figure}[!ht]
\centering
\includegraphics[width=7.5cm, clip]{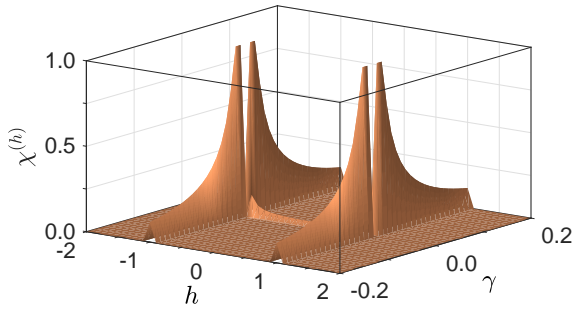}\\
\caption{(Color online) Normalized fidelity susceptibility $\chi^{(h)}$ with respect to $h$. 
It has a sharp extremum at the critical lines $|h|=1$
where continuous QPTs with $\nu=1$ occur.}\label{FIG-FSVh}
\end{figure}

We now consider the FS at some special values.
When $\gamma=1$, the $XY$ model is reduced to the TFIM, and the FS has the form
\begin{align}\label{FSTFIM}
&\chi^{(h)}(h,1) = \frac{\mathcal{N}}{16h^2}\cdot \nonumber \\
&\qquad \left[ \frac{\mathcal{N}h^{\mathcal{N}}}{(h^{\mathcal{N}}+1)^2} + \frac12\left(\frac{(h^2+1)(h^{\mathcal{N}}-1)}{(h^2-1)(h^{\mathcal{N}}+1)}-1\right) \right],
\end{align}
which agrees with the result by finite sums of hyperbolic functions\cite{Damski2013}.
When $h=h_{c}$, i.e., $|h|=1$, the FS is
\begin{align}\label{FShEQhc}
&\chi^{(h)}(h_c,\gamma) = \frac{\mathcal{N}^2-\frac{3-\gamma^2}{2\gamma}\mathcal{N}}{32\gamma^2} \nonumber \\
&\qquad +\frac{\mathcal{N}\left[\Big(\mathcal{N}+\frac{3-\gamma^2}{2\gamma}\Big)\big(\frac{1+\gamma}{1-\gamma}\big)^{\mathcal{N}} + \frac{3-\gamma^2}{2\gamma}\right] }{16\gamma^2\Big[\big(\frac{1+\gamma}{1-\gamma}\big)^{\mathcal{N}}+1\Big]^2}.
\end{align}
The first term of Eq.~\eqref{FShEQhc} is the leading term,
while the second term is either exponential decay
or linearly increased with the system size $\mathcal{N}$, depending on the sign of $\gamma$.
In the finite-size system, the maximal value of the FS $\chi^{(h)}(h=h_m,\gamma)$
is slightly larger than that of $\chi^{(h)}(h=h_c,\gamma)$,
but with the quadratic term of $\mathcal{N}$ unchanged. Therefore,
$\chi^{(h)}(h_m,\gamma) \simeq \chi^{(h)}(h_c,\gamma)
\simeq \frac{\mathcal{N}}{32\gamma^2}\big(\mathcal{N}-\frac{3-\gamma^2}{2|\gamma|}\big)$.
Considering only the leading term, we have
\begin{equation}\label{FShMaxFSS}
\ln\left(\frac{\chi^{(h)}(h_m,\gamma)}{\mathcal{N}}\right) = \ln\mathcal{N}-\ln(32\gamma^2).
\end{equation}
In the TDL, the exact expression for the FS can be calculated by the residue theorem and turns out to be\cite{Kolodrubetz2013}
\begin{align}\label{FShTDL}
\frac{\bar{\chi}^{(h)}(h,\gamma)}{\mathcal{N}} = \frac{1}{16}
\left\{
  \begin{array}{ll}
    \vspace{0.1cm}
    \frac{1}{|\gamma|(1-h^2)}, & |h|<1 \\
    \frac{|h|\gamma^2}{(h^2-1)(h^2+\gamma^2-1)^{3/2}}, & |h|>1
  \end{array}
\right.,
\end{align}
and the scaling behavior around the critical points is
\begin{equation}\label{FShTDLFSS}
\ln\left(\frac{\bar{\chi}^{(h)}(h,\gamma)}{\mathcal{N}}\right) = -\ln|h-h_c|-\ln(32\gamma).
\end{equation}
It can be noticed that the absolute value of the prefactors of Eq.~\eqref{FShMaxFSS} and Eq.~\eqref{FShTDLFSS} are equal,
indicating that the critical component $\nu=1$ according to the ansatz Eq.~\eqref{ChiScale}.
The normalized FS $\chi^{(h)}(h,\gamma)$ with respect to $h$ in the TDL is shown in FIG.~\ref{FIG-FSVh}.
It is found that the FS has a sharp extremum at $|h|=1$ where a continuous QPT occurs.
This results demonstrate convincingly that FS can be used to characterize the quantum critical behavior.

\subsection{The anisotropy transition}
\begin{figure}[!ht]
\centering
\includegraphics[width=7.5cm, clip]{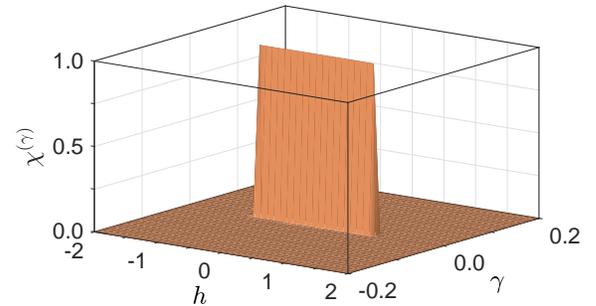}\\
\caption{(Color online) Normalized fidelity susceptibility $\chi^{(\gamma)}$ with respect to $\gamma$. 
It has a sharp extremum at the segment $|h|<1$, $\gamma=0$
where continuous QPTs with $\nu=1$ occur.}\label{FIG-FSVr}
\end{figure}

Calculation for the FS with respect to the anisotropy transition is similar to
the Ising transition discussed above.
To begin with, we should rewrite the expression of Eq.~\eqref{DefXYFSr} as
\begin{align}\label{DefXYFSrv1}
\chi^{(\gamma)}(h,\gamma)
&= \frac{1}{4(1-\gamma^2)^2}\sum_{n=1}^{N}\frac{(1-c_n^2)(c_n-h)^2}{(c_n-\lambda_{+})^2(c_n-\lambda_{-})^2}
\end{align}
where $\lambda_{\upsilon}$'s~($\upsilon=\pm$) are defined in Eq.~\eqref{lambdaNu}.
Once the partial fraction expansion of the expression $\frac{(1-c_n^2)(c_n-h)^2}{(c_n-\lambda_{+})^2(c_n-\lambda_{-})^2}$
is obtained, the FS is readily splitted into two symmetric forms and
each of them could be calculated according to the method shown in the last subsection.
So we neglect the tedious details and present the final result
\begin{equation}\label{FS-XYWRTr}
\chi^{(\gamma)}(h,\gamma) = \frac{1}{16(h^2+\gamma^2-1)}\sum_{\upsilon=\pm}\mathcal{F}_{\upsilon}^{(\gamma)}-\frac{\mathcal{N}}{8(1-\gamma^2)^2}
\end{equation}
where
\begin{equation}\label{FsumWRTr}
\mathcal{F}_{\upsilon}^{(\gamma)} =\frac{(g_{\upsilon}^2-1)^2}{4g_{\upsilon}^2}
\left[\frac{\mathcal{N}^2g_{\upsilon}^{\mathcal{N}}}{(g_{\upsilon}^{\mathcal{N}}+1)^2} + \mathcal{N}\Big(\frac{C_{\upsilon}}{2}+\frac{p_{\upsilon}}{|\gamma|}\Big)\frac{g_{\upsilon}^{\mathcal{N}}-1}{g_{\upsilon}^{\mathcal{N}}+1}\right].
\end{equation}
Here, $p_{\upsilon}$, $g_{\upsilon}$, and $C_{\upsilon}$ are as defined earlier.
Again, the detailed derivations are presented in the SM\cite{MySUPP}.
It is well-known that the anisotropy transition occurs when $|h|<1$,
and in this segment $\chi^{(\gamma)}(h,\gamma)$ reaches its maximum \textit{exactly} at $\gamma=0$.
Precisely, we have
\begin{equation}\label{FS-XYWRTrEQ0}
\chi^{(\gamma)}(h,0) = -\frac{\mathcal{N}}{8} + \frac{1}{4}\left[\frac{\mathcal{N}^2g^{\mathcal{N}}}{(g^{\mathcal{N}}+1)^2} + \frac{\mathcal{N}h}{2i\sqrt{1-h^2}}\frac{g^{\mathcal{N}}-1}{g^{\mathcal{N}}+1}\right]
\end{equation}
where $g=h+i\sqrt{1-h^2}$.
Physically, $\chi^{(\gamma)}(h,0)$ is a real expression with vanishing imaginary part.
This can be seen clearly by parameterizing $h=\cos\theta$,~$\theta\in(0,\pi)$,
\begin{align}\label{FS-XYWRTrEQ0v2}
&\chi^{(\gamma)}(h,0) = \nonumber\\
&\quad \frac{N}{4}\left[ \frac{N}{\cos^2(N\cos^{-1}(h))} + \frac{\tan(N\cos^{-1}(h))}{\tan(\cos^{-1}(h))} -1 \right].
\end{align}
It can be proved that the ratio of the second term to the first one in the bracket
is an oscillating function and is bounded.
So the second term can be neglected in that it contributes the leading term to a prefactor at most.
The third term also does not need to be considered for large enough system size $\mathcal{N}$.
In light of the relation
$\chi^{(\gamma)}(h,\gamma_m)$ = $\chi^{(\gamma)}(h,\gamma_c)$ ~ $\simeq$
$\frac{{\mathcal{N}}^2}{16\cos^2(\mathcal{N}\cos^{-1}(h)/2)}$,
we obtain that
\begin{equation}\label{FSrMaxFSS}
\ln\left(\frac{\chi^{(\gamma)}(h,0)}{\mathcal{N}}\right) = \ln\mathcal{N}-\ln\left[16\cos^2\frac{\mathcal{N}\cos^{-1}(h)}{2}\right].
\end{equation}
Similarly, the exact expression for the FS is\cite{Kolodrubetz2013}
\begin{align}\label{FSrTDL}
\frac{\bar{\chi}^{(r)}(h,\gamma)}{\mathcal{N}} = \frac{1}{16}\frac{1}{|\gamma|(1+|\gamma|)^2},\qquad |h|<1
\end{align}
in the TDL, indicating that the scaling behavior around the critical points is
\begin{equation}\label{FSrTDLFSS}
\ln\left(\frac{\bar{\chi}^{(\gamma)}(h,\gamma)}{\mathcal{N}}\right) = -\ln|\gamma|-4\ln2.
\end{equation}
Similarly, we are safe to conclude from Eq.~\eqref{FSrMaxFSS} and Eq.~\eqref{FSrTDLFSS}
that the critical exponent $\nu=1$ according to the ansatz Eq.~\eqref{ChiScale}.
Besides, the normalized FS $\chi^{(\gamma)}(h,\gamma)$ with respect to $\gamma$ in the TDL is shown in FIG.~\ref{FIG-FSVr}.
This results also demonstrate convincingly that FS can be used to characterize the quantum critical behavior.
\section{Conclusion}\label{SecConc}
In conclusion, we have derived several closed-form expressions for fidelity susceptibility~(FS) of the
anisotropic $XY$ model in the transverse field after a symmetry analysis.
The FS at the special case $\gamma=1$ can be recovered and is consistent with the results obtained by
Damski \textit{et~al.} in light of several elegant summation formulas\cite{Damski2013,Damski2014,Damski2015}
on the transverse-field Ising model.
Our method is easy to follow and promises to be useful to other exactly solvable models,
such as the $XY$ model with bond-alternated interaction\cite{Taylor1985} or staggered field\cite{Perk1975}.
The correlation length critical exponent $\nu=1$ is calculated analytically according to the
standard finite-size scaling ansatz.

\textit{Acknowledgments}--We thank W.-L. You for some useful discussions.
Q.L. was financially supported by the Outstanding Innovative Talents
Cultivation Funded Programs 2017 of Renmin University of China.
J.Z. was supported by the the National Natural Science Foundation of China (Grant No. 11474029).
X.W. was supported by the National Program on Key Research Project (Grant No. 2016YFA0300501)
and by the National Natural Science Foundation of China (Grant No. 11574200).

\appendix
\setcounter{figure}{0}
\renewcommand{\thefigure}{A.\arabic{figure}}
\setcounter{equation}{0}
\renewcommand{\theequation}{A.\arabic{equation}}
\section{Closed-Form Expression}\label{AppA}
\textit{Lemma 1}~For any nonzero (complex) variable $\alpha$,
the closed-form of the summation
\begin{equation}\label{Deffalpha}
L(\alpha) = \sum_{n=1}^{N} \frac{1}{\alpha+c_n},\;c_n = \cos\frac{(2n-1)\pi}{\mathcal{N}}
\end{equation}
owns the following expression:
\begin{align}\label{ClosedFormFalpha}
L(\alpha)
&=
\left\{
  \begin{array}{ll}
    \vspace{0.1cm}
    \frac{N}{\sqrt{\alpha^2-1}} \frac{\beta^{\mathcal{N}}-1}{\beta^{\mathcal{N}}+1},   & \vert\alpha\vert > 1 \\
    \vspace{0.1cm}
    \textrm{sgn}(\alpha) N^2,   & \vert\alpha\vert \to 1  \\
    \vspace{0.1cm}
    \frac{N}{\sqrt{1-\alpha^2}} \tan(N \cos^{-1}(\alpha)),   & \vert\alpha\vert < 1
  \end{array}
\right..
\end{align}
where $\mathcal{N}=2N$, $\beta = \alpha + \sqrt{\alpha^2-1}$ and $\textrm{sgn}(\cdot)$ is the sign function.
\begin{figure}[!ht]
\centering
\includegraphics[width=7.5cm, clip]{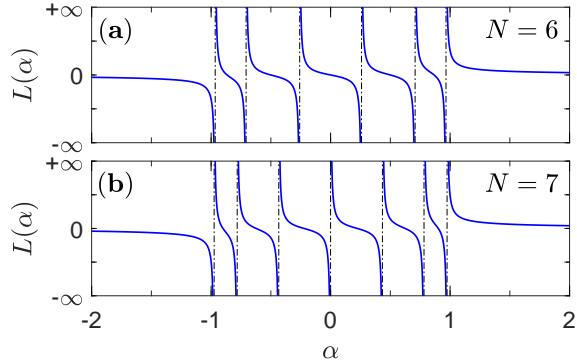}\\
\caption{(Color online) Illustration of the function $L(\alpha)$ for
  (a)~the even case $N=6$ and for (b)~the odd case $N=7$.
  In both cases, the solid~(blue) lines are the curve for $L(\alpha)$,
  while the dash-dotted~(black) lines represent the discontinuous points.}\label{FIG-FuncAlpha}
\end{figure}

\begin{proof}
The function $L(\alpha)$ is an odd function with respect to the variable $\alpha$,
which can be verified immediately by noticing the symmetry relation $c_n = -c_{N+1-n}$.
The monotonic behavior of the function depends highly on $\alpha$ and
we shall consider the case $\vert\alpha\vert > 1$ firstly.

To begin with, let $\zeta=e^{\frac{\pi}{\mathcal{N}}i}$ be the primitive $2\mathcal{N}$-th root of unity and
\begin{equation*}
\Omega = \{\zeta^{2n-1}~\vert -N<n\leq N\} = \{\zeta^{\pm(2n-1)}~\vert 1\leq n\leq N\}
\end{equation*}
be the set of roots for the polynomial $z^{\mathcal{N}}+1=0$.
Therefore, for any allowed integer $n$,
$c_n = \frac{\omega+\omega^{-1}}{2}$ with $\omega$ = $\omega_n$ = $\zeta^{2n-1}$.
Employing the logarithmic derivative with respect to $z$ of the equality
\begin{equation*}
\prod_{\omega\in\Omega}(z+\omega) = \prod_{\omega\in\Omega}(z-\omega) = z^{\mathcal{N}}+1,
\end{equation*}
we then arrive at the important summation identity
\begin{equation}\label{EqualityZoverOmega}
\sum_{\omega\in\Omega}\frac{z}{\omega+z} = \mathcal{N}\frac{z^{\mathcal{N}}}{z^{\mathcal{N}}+1}.
\end{equation}

Actually, it is natural for us to define the new variable $\beta=\alpha+\sqrt{\alpha^2-1}$
so that $\alpha$ = $\frac{\beta+\beta^{-1}}{2}$. 
In light of Eq.~\eqref{EqualityZoverOmega}, we have
\begin{align}\label{DeriveFalphainBeta}
L(\alpha)
&= \frac{1}{2}\sum_{n=-N+1}^{N} \frac{1}{\alpha+c_n} 
= \frac{1}{2}\sum_{\omega\in\Omega} \frac{1}{\frac{\beta+\beta^{-1}}{2} + \frac{\omega+\omega^{-1}}{2}}  \nonumber\\
&= \frac{1}{\beta-\beta^{-1}}\sum_{\omega\in\Omega} \left(\frac{\beta}{\omega+\beta} - \frac{\beta^{-1}}{\omega+\beta^{-1}} \right) \nonumber\\
&= \frac{N}{\sqrt{\alpha^2-1}}\frac{\beta^{\mathcal{N}}-1}{\beta^{\mathcal{N}}+1}.
\end{align}
If the hyperbolic functions are involved, Eq.~\eqref{DeriveFalphainBeta} is equivalent to
\begin{align}\label{DeriveFalphainHyperBolic}
L(\alpha) &= \frac{N\cdot\textrm{sgn}(\alpha)}{\sqrt{\alpha^2-1}} \tanh(N \cosh^{-1}(\alpha)).
\end{align}

The exact expression for the case $|\alpha|<1$ can be obtained from the former
in the spirit of analytic continuation or simply
by new substitution $\beta = \alpha + i\sqrt{1-\alpha^2}$.
In this situation, we have
\begin{align}\label{DeriveFalphainAlphaLT1}
L(\alpha) &= \frac{N}{\sqrt{1-\alpha^2}} \tan(N \cos^{-1}(\alpha)).
\end{align}
It should be note that the point $\alpha=0$ is out of our consideration in general
since whether this point make sense or not depends on the odevity of $N$.
It is zero when $N$ is even and is diverging for odd case, see Fig.~\ref{FIG-FuncAlpha}.
More generally, there are $N$ discontinuous points in the range of $|\alpha|<1$; namely,
$\alpha_{\textrm{disc}}$=$\cos\frac{2m+1}{\mathcal{N}}\pi$, $m$=0,1,2,$\cdots$,$N-1$.
Therefore, the function $L(\alpha)$ is continuous at the points $|\alpha|=1$ and
its values thereof are equal to $\textrm{sgn}(\alpha)\cdot N^2$.

All together, we finish the full processes of the proof.
We end this section by the Fig.~\ref{FIG-FuncAlpha} which presents the curvature of function $L(\alpha)$
for even and odd $N$.
\end{proof}

\setcounter{figure}{0}
\renewcommand{\thefigure}{B.\arabic{figure}}
\setcounter{equation}{0}
\renewcommand{\theequation}{B.\arabic{equation}}
\section{Derivative Relation}\label{AppB}
\textit{Lemma 2} For any nonzero (complex) $\alpha$, the first derivative of the function $L(\alpha)$
defined in Eq.~\eqref{Deffalpha} is
\begin{align}\label{ClosedFormDiffFalpha}
&L'(\alpha)= \frac{\partial L(\alpha)}{\partial \alpha}  \nonumber\\
&=
\left\{
  \begin{array}{ll}
    \vspace{0.1cm}
    \frac{4\beta^2}{(\beta^2-1)^2}
    \left[\frac{\mathcal{N}^2\beta^{\mathcal{N}}}{(\beta^{\mathcal{N}}+1)^2}
    - \frac{\mathcal{N}}{2}\frac{(\beta^2+1)(\beta^{\mathcal{N}}-1)}{(\beta^2-1)(\beta^{\mathcal{N}}+1)}\right],   & \vert\alpha\vert > 1 \\
     \vspace{0.1cm}
    -\textrm{sgn}(\alpha)\frac{N^2(16N^2+9)}{24},                           & \vert\alpha\vert \to 1  \\
    -\frac{L(\alpha)}{\sqrt{1-\alpha^2}}
    \left[{\frac{\mathcal{N}}{\sin(\mathcal{N}\cos^{-1}(\alpha))}
    - \frac{\alpha}{\sqrt{1-\alpha^2}}}\right],                             & \vert\alpha\vert < 1
  \end{array}
\right.
\end{align}
where $\mathcal{N}=2N$, $\beta = \alpha + \sqrt{\alpha^2-1}$ and $\textrm{sgn}(\cdot)$ is the sign function.

\setcounter{figure}{0}
\renewcommand{\thefigure}{C.\arabic{figure}}
\setcounter{equation}{0}
\renewcommand{\theequation}{C.\arabic{equation}}
\section{Quantum Metric Tensor}\label{AppC}
The quantum metric tensor\cite{Zanardi2007} is a concept stemming from differential geometry
and information theory. It describes the absolute value of the overlap amplitude
between neighboring ground states.
Therefore, like the fidelity susceptibility, The metric also plays an vital role
in understanding the quantum phase transition.

For the anisotropic $XY$ model~(see Eq.~\eqref{XYModel}) in the ($h$,$\gamma$) parameter space,
the quantum metric tensor is defined as
$g_{\sigma\bar\sigma} = \frac{1}{4}\sum_{n=1}^{N}
({\partial \theta_k}/{\partial \mu^{\sigma}})({\partial \theta_k}/{\partial \mu^{\bar\sigma}})$,
where $\mu^{1,2} = h$, $\gamma$\cite{Zanardi2007}.
So, the solely off-diagonal element of the tensor is
\begin{align}\label{DefXYFShr}
\chi^{(h\gamma)}(h,\gamma) &= \frac{\gamma}{4}\sum_{k>0}\frac{\sin^2k(\cos k-h)}{\Big[(\cos k-h)^2+\gamma^2\sin^2k\Big]^2}.
\end{align}

Method for the calculation of Eq.~\eqref{DefXYFShr}
has been explained in the main text. The detail is omitted here
and is presented in the SM\cite{MySUPP}, and the result is
\begin{align}\label{XYFSOffDiagCHInu}
\chi^{(h\gamma)}(h,\gamma)
=\frac{\textrm{sgn}(\gamma)}{16(h^2+\gamma^2-1)}\sum_{\upsilon=\pm}\mathcal{F}_{\upsilon}^{(h\gamma)}
\end{align}
where
\begin{align}\label{XYFSOffDiagFnu}
\mathcal{F}_{\upsilon}^{(h\gamma)}
&= \frac{g_{\upsilon}^2-1}{2p_{\upsilon}g_{\upsilon}}
\Bigg[ \frac{\mathcal{N}^2g_{\upsilon}^{\mathcal{N}}}{(g_{\upsilon}^{\mathcal{N}}+1)^2} + \frac{\mathcal{N}}{2}
\left(\frac{p_{\upsilon}}{|\gamma|}+C_{\upsilon}\right)
\frac{g_{\upsilon}^{\mathcal{N}}-1}{g_{\upsilon}^{\mathcal{N}}+1} \Bigg].
\end{align}
Also, the $g_{\upsilon}$, $p_{\upsilon}$, and $C_{\upsilon}$ are defined in
Eq.~\eqref{Def:gbeta}, Eq.~\eqref{Def:qsgn}, and Eq.~\eqref{Def:Coeff}.


\bibliography{manuscript}





\clearpage

\onecolumngrid

\newpage

\newcounter{equationSM}
\newcounter{figureSM}
\newcounter{tableSM}
\stepcounter{equationSM}
\setcounter{equation}{0}
\setcounter{figure}{0}
\setcounter{table}{0}
\setcounter{page}{1}
\makeatletter
\renewcommand{\theequation}{\textsc{sm}-\arabic{equation}}
\renewcommand{\thefigure}{\textsc{sm}-\arabic{figure}}
\renewcommand{\thetable}{\textsc{sm}-\arabic{table}}

\onecolumngrid

\begin{center}
{\large{\bf Supplemental Material for\\ ``Fidelity susceptibility of the anisotropic $XY$ model: The exact solution''}}
\end{center}
\begin{center}
Qiang Luo$^{1, *}$, Jize Zhao$^{2}$, and Xiaoqun Wang$^{3, 4, \dagger}$\\
\quad\\
$^1$\textit{Department of Physics, Renmin University of China, Beijing 100872, China}\\
$^2$\textit{Center for Interdisciplinary Studies, Lanzhou University, Lanzhou 730000, China}\\
$^3$\textit{Key Laboratory of Artificial Structures and Quantum Control (Ministry of Education), School of Physics and Astronomy, Tsung-Dao Lee Institute, Shanghai Jiao Tong University, Shanghai 200240, China}\\
$^4$\textit{Collaborative Innovation Center for Advanced Microstructures, Nanjing 210093, China}
\end{center}
In this supplemental material~(SM), we collect a detailed derivation of the fidelity susceptibility~(FS) presented in the main text.
For simplicity, we will still assume
$|\gamma|<1$ and $h^2+\gamma^2>1$ so as to avoid the imaginary unit $i$.

\subsection{Basic Propositions}
Let us recall that $\phi(t)$=$t^2-\frac{2h}{1-\gamma^2}t + \frac{h^2+\gamma^2}{1-\gamma^2}$=$(t-\lambda_{+})(t-\lambda_{-})$,
and the \textit{real} roots of the equation $\phi(t)=0$ read
\begin{align}\label{SMlambdaNu}
\lambda_{\upsilon} = \frac{h + \upsilon\vert\gamma\vert\sqrt{h^2+\gamma^2-1}}{1-\gamma^2}.
\end{align}
where $\upsilon=\pm$. Below we will give four relevant propositions which will be used frequently in the SM.

\textit{Proposition 1.}--
Let $g_{\upsilon} = \lambda_{\upsilon}+\sqrt{\lambda_{\upsilon}^2-1}$, then we have
\begin{align}\label{SMDef:LmbSqrtLmb}
\lambda_{\upsilon} = \frac{g_{\upsilon}^2+1}{2g_{\upsilon}},\;
\sqrt{\lambda_{\upsilon}^2-1} = \frac{g_{\upsilon}^2-1}{2g_{\upsilon}},\;
\frac{\lambda_{\upsilon}}{\sqrt{\lambda_{\upsilon}^2-1}} = \frac{g_{\upsilon}^2+1}{g_{\upsilon}^2-1}.
\end{align}

\textit{Proposition 2.}--
The following identity holds:
\begin{align}\label{SMDef:Kmb-Fldh}
\lambda_{\upsilon}-h = \frac{|\gamma|}{p_{\upsilon}}\sqrt{\lambda_{\upsilon}^2-1}
\end{align}
where $p_{\upsilon}$=$p_{\upsilon}(h)$ is a piecewise sign function of field $h$ and branch $\upsilon$
\begin{align}\label{SMDef:qsgn}
p_{\upsilon} =
\left\{
  \begin{array}{ll}
    \upsilon, & |h|>1 \\
    \mathrm{sgn}(h), & |h|<1
  \end{array}
\right..
\end{align}

\textit{Proposition 3.}--
Let $g_{\upsilon} = \lambda_{\upsilon}+\sqrt{\lambda_{\upsilon}^2-1}$, then we have
\begin{align}\label{SMDef:gbeta}
g_{\upsilon} = \frac{h+\upsilon p_{\upsilon}\sqrt{h^2+\gamma^2-1}}{1-p_{\upsilon}|\gamma|}
\end{align}

\begin{proof}
To begin with, when $|\gamma|<1$ we have
\begin{align*}
\sqrt{\lambda_{\upsilon}^2-1}
=\sqrt{ \left(\frac{h + \upsilon\vert\gamma\vert\sqrt{h^2+\gamma^2-1}}{1-\gamma^2}\right)^2-1}
=\frac{\big\vert h\vert\gamma\vert + \upsilon\sqrt{h^2+\gamma^2-1} \big\vert}{1-\gamma^2}.
\end{align*}
The absolute value sign should be taken out to further simplify the expression. This can be done by noticing that
\begin{equation*}
\big(h\vert\gamma\vert + \sqrt{h^2+\gamma^2-1}\big)\big(h\vert\gamma\vert - \sqrt{h^2+\gamma^2-1}\big)
=(1-h^2)(1-\gamma^2).
\end{equation*}
That's means, the sign of the expression in the absolute value is equal to the sign of field $h$ when $|h|<1$ or equal to the sign of branch $\upsilon$ when $|h|>1$. So,
\begin{align*}
\sqrt{\lambda_{\upsilon}^2-1}
=\frac{p_{\upsilon}\big( h\vert\gamma\vert + \upsilon\sqrt{h^2+\gamma^2-1} \big)}{1-\gamma^2}
\end{align*}
where $p_{\upsilon}$ is defined in \eqref{SMDef:qsgn}. In this case we have
\begin{align*}
g_{\upsilon} &= \lambda_{\upsilon}+\sqrt{\lambda_{\upsilon}^2-1} \\
&= \frac{h + \upsilon\vert\gamma\vert\sqrt{h^2+\gamma^2-1}}{1-\gamma^2}
+\frac{p_{\upsilon}\big( h\vert\gamma\vert + \upsilon\sqrt{h^2+\gamma^2-1} \big)}{1-\gamma^2} \\
&= \frac{(1+p_{\upsilon}\vert\gamma\vert)h + \upsilon(\vert\gamma\vert+p_{\upsilon})\sqrt{h^2+\gamma^2-1}}{1-\gamma^2} \\
&= \frac{(1+p_{\upsilon}\vert\gamma\vert)h + \upsilon p_{\upsilon}(p_{\upsilon}\vert\gamma\vert+1)\sqrt{h^2+\gamma^2-1}}{1-\gamma^2} \\
&= \frac{h+\upsilon p_{\upsilon}\sqrt{h^2+\gamma^2-1}}{1-p_{\upsilon}|\gamma|}
\end{align*}
where in the penultimate step we utilize the property of sign function, i.e., $p_{\upsilon}^2=1$.
\end{proof}

\textit{Proposition 4.}--
Let $C_{\upsilon} = \frac{2-\lambda_{\upsilon}\lambda_{\bar\upsilon}-\lambda_{\upsilon}^2}
{(\lambda_{\upsilon}-\lambda_{\bar\upsilon})\sqrt{\lambda_{\upsilon}^2-1}}$
where $\bar\upsilon$ is the complementary component of branch $\upsilon$~($\bar\upsilon\upsilon=-1$),
then we have
\begin{align}\label{SMDef:Coeff}
C_{\upsilon} =\frac{p_{\upsilon}}{h^2-1}
\left[\frac{\upsilon{\gamma}^2h}{\sqrt{h^2+\gamma^2-1}}-\frac{h^2+\gamma^2-1}{|\gamma|}\right].
\end{align}

\begin{proof}
We now firstly calculate the numerator of $C_{\upsilon}$. In light of Eq.~\eqref{SMlambdaNu}, we have
\begin{align*}
2-\lambda_{\upsilon}\lambda_{\bar\upsilon}-\lambda_{\upsilon}^2
&= 2 - \frac{h^2+\gamma^2}{1-\gamma^2} - \left(\frac{h + \upsilon\vert\gamma\vert\sqrt{h^2+\gamma^2-1}}{1-\gamma^2}\right)^2  \\
&= \frac{2(1-\gamma^2)^2 - [h^2-\gamma^2(h^2+\gamma^2-1)] - [h^2+\gamma^2(h^2+\gamma^2-1) + 2\upsilon h\vert\gamma\vert\sqrt{h^2+\gamma^2-1}]}
{(1-\gamma^2)^2}  \\
&= \frac{2[(1-\gamma^2)^2 - h^2] - 2\upsilon h\vert\gamma\vert\sqrt{h^2+\gamma^2-1}} {(1-\gamma^2)^2}.
\end{align*}
Consequently, the expression $C_{\upsilon}$ turns out to be
\begin{align*}
C_{\upsilon}
&= \frac{2-\lambda_{\upsilon}\lambda_{\bar\upsilon}-\lambda_{\upsilon}^2}
{(\lambda_{\upsilon}-\lambda_{\bar\upsilon})\sqrt{\lambda_{\upsilon}^2-1}} \\
&= \frac{2[(1-\gamma^2)^2 - h^2] - 2\upsilon h\vert\gamma\vert\sqrt{h^2+\gamma^2-1}} {(1-\gamma^2)^2}\cdot
\frac{1-\gamma^2}{2\upsilon\vert\gamma\vert\sqrt{h^2+\gamma^2-1}}\cdot
\frac{1-\gamma^2}{p_{\upsilon}\big( h\vert\gamma\vert + \upsilon\sqrt{h^2+\gamma^2-1} \big)}  \\
&= \frac{[(1-\gamma^2)^2 - h^2] - \upsilon h\vert\gamma\vert\sqrt{h^2+\gamma^2-1}}
{\upsilon p_{\upsilon}\vert\gamma\vert \sqrt{h^2+\gamma^2-1} \big( h\vert\gamma\vert + \upsilon\sqrt{h^2+\gamma^2-1} \big)} \\
&= \frac{\Big\{[(1-\gamma^2)^2 - h^2] - \upsilon h\vert\gamma\vert\sqrt{h^2+\gamma^2-1}\Big\} \big( h\vert\gamma\vert - \upsilon\sqrt{h^2+\gamma^2-1} \big)}
{\upsilon p_{\upsilon}\vert\gamma\vert \sqrt{h^2+\gamma^2-1} \big( h\vert\gamma\vert + \upsilon\sqrt{h^2+\gamma^2-1} \big)\big( h\vert\gamma\vert - \upsilon\sqrt{h^2+\gamma^2-1} \big)} \\
&= \frac{\gamma^2(\gamma^2-1) h\vert\gamma\vert - \upsilon(\gamma^2-1)(h^2+\gamma^2-1)\sqrt{h^2+\gamma^2-1}}
{\upsilon p_{\upsilon}\vert\gamma\vert \sqrt{h^2+\gamma^2-1} (h^2-1)(\gamma^2-1)} \\
&= \frac{\gamma^2 h\vert\gamma\vert - \upsilon(h^2+\gamma^2-1)\sqrt{h^2+\gamma^2-1}}
{\upsilon p_{\upsilon}\vert\gamma\vert \sqrt{h^2+\gamma^2-1} (h^2-1)} \\
&= \frac{p_{\upsilon}}{h^2-1}
\left[\frac{\upsilon{\gamma}^2h}{\sqrt{h^2+\gamma^2-1}}-\frac{h^2+\gamma^2-1}{|\gamma|}\right].
\end{align*}
\end{proof}

\subsection{Calculation for the Ising transition}
The formula for the FS of the Ising transition is
\begin{align}\label{SMDefXYFShv1}
\chi^{(h)}(h,\gamma)
&= \frac{\gamma^2}{4(1-\gamma^2)^2}\sum_{n=1}^{N}\frac{1-c_n^2}{(c_n-\lambda_{+})^2(c_n-\lambda_{-})^2}
\end{align}
where $c_n = \cos\frac{(2n-1)\pi}{\mathcal{N}}$~($\mathcal{N}=2N$) and
$\lambda_{\upsilon}$'s~($\upsilon=\pm$) are defined in Eq.~\eqref{SMlambdaNu}.
To eliminate the quadratic terms, we shall utilize the partial fraction expansion of the expression $\frac{1-c_n^2}{(c_n-\lambda_{+})^2(c_n-\lambda_{-})^2}$.
It is known that for any $a$ and $b$, we have
\begin{equation*}
\frac{1-x^2}{(x-a)^2(x-b)^2} = \mathcal{F}(a,b) + \mathcal{F}(b,a)
\end{equation*}
where
\begin{equation*}
\mathcal{F}(a,b) = \frac{1-a^2}{(a-b)^2(x-a)^2} + \frac{2(ab-1)}{(a-b)^3(x-a)}.
\end{equation*}
Therefore, Eq.~\eqref{SMDefXYFShv1} can be recast into a symmetric form by $\mathcal{F}$.

We now simplify the summation
$\mathcal{F}_{\upsilon}^{(h)} = (\lambda_{\upsilon}-\lambda_{\bar\upsilon})^2
\sum_{n=1}^{N}\mathcal{F}(\lambda_{\upsilon},\lambda_{\bar\upsilon})$,
which reads,
\begin{align*}
\mathcal{F}_{\upsilon}^{(h)}
&= \sum\frac{1-\lambda_{\upsilon}^2}{(c_n-\lambda_{\upsilon})^2} + \sum\frac{2(\lambda_{\upsilon}\lambda_{\bar\upsilon}-1)}{(\lambda_{\upsilon}-\lambda_{\bar\upsilon})(c_n-\lambda_{\upsilon})} \nonumber\\
&= (\lambda_{\upsilon}^2-1)L'(-\lambda_{\upsilon}) + \frac{2(\lambda_{\upsilon}\lambda_{\bar\upsilon}-1)}{\lambda_{\upsilon}-\lambda_{\bar\upsilon}}L(-\lambda_{\upsilon}) \nonumber\\
&= (\lambda_{\upsilon}^2-1)L'(\lambda_{\upsilon}) - \frac{2(\lambda_{\upsilon}\lambda_{\bar\upsilon}-1)}{\lambda_{\upsilon}-\lambda_{\bar\upsilon}}L(\lambda_{\upsilon})
\end{align*}
where in the last step we have used the odevity of function $L(\alpha)$ and its derivative.
By virtue of the two Lemmas listed in the APPENDIX of the main text, we have
\begin{align*}
\mathcal{F}_{\upsilon}^{(h)}
&= (\lambda_{\upsilon}^2-1)
\frac{4g_{\upsilon}^2}{(g_{\upsilon}^2-1)^2}
    \left[\frac{\mathcal{N}^2g_{\upsilon}^{\mathcal{N}}}{(g_{\upsilon}^{\mathcal{N}}+1)^2}
    - \frac{\mathcal{N}}{2}
    \frac{(g_{\upsilon}^2+1)(g_{\upsilon}^{\mathcal{N}}-1)}{(g_{\upsilon}^2-1)(g_{\upsilon}^{\mathcal{N}}+1)}\right]
-\frac{2(\lambda_{\upsilon}\lambda_{\bar\upsilon}-1)}{\lambda_{\upsilon}-\lambda_{\bar\upsilon}}
 \frac{\mathcal{N}}{2\sqrt{\lambda_{\upsilon}^2-1}} \frac{g_{\upsilon}^{\mathcal{N}}-1}{g_{\upsilon}^{\mathcal{N}}+1} \\
&= \frac{\mathcal{N}^2g_{\upsilon}^{\mathcal{N}}}{(g_{\upsilon}^{\mathcal{N}}+1)^2} + \frac{\mathcal{N}}{2}
\left[-\frac{g_{\upsilon}^2+1}{g_{\upsilon}^2-1}
-\frac{2(\lambda_{\upsilon}\lambda_{\bar\upsilon}-1)}{(\lambda_{\upsilon}-\lambda_{\bar\upsilon})\sqrt{\lambda_{\upsilon}^2-1}}\right]
\frac{g_{\upsilon}^{\mathcal{N}}-1}{g_{\upsilon}^{\mathcal{N}}+1}  \\
&= \frac{\mathcal{N}^2g_{\upsilon}^{\mathcal{N}}}{(g_{\upsilon}^{\mathcal{N}}+1)^2} + \frac{\mathcal{N}}{2}
\left[-\frac{\lambda_{\upsilon}}{\sqrt{\lambda_{\upsilon}^2-1}}
-\frac{2(\lambda_{\upsilon}\lambda_{\bar\upsilon}-1)}{(\lambda_{\upsilon}-\lambda_{\bar\upsilon})\sqrt{\lambda_{\upsilon}^2-1}}\right]
\frac{g_{\upsilon}^{\mathcal{N}}-1}{g_{\upsilon}^{\mathcal{N}}+1}  \\
&= \frac{\mathcal{N}^2g_{\upsilon}^{\mathcal{N}}}{(g_{\upsilon}^{\mathcal{N}}+1)^2} + \frac{\mathcal{N}}{2}
\left[\frac{2-\lambda_{\upsilon}\lambda_{\bar\upsilon}-\lambda_{\upsilon}^2}
{(\lambda_{\upsilon}-\lambda_{\bar\upsilon})\sqrt{\lambda_{\upsilon}^2-1}}\right]
\frac{g_{\upsilon}^{\mathcal{N}}-1}{g_{\upsilon}^{\mathcal{N}}+1}  \\
&= \frac{\mathcal{N}^2g_{\upsilon}^{\mathcal{N}}}{(g_{\upsilon}^{\mathcal{N}}+1)^2} + \frac{\mathcal{N}C_{\upsilon}}{2}\frac{g_{\upsilon}^{\mathcal{N}}-1}{g_{\upsilon}^{\mathcal{N}}+1}
\end{align*}
Here, \textit{Proposition 1} is used in the first step and is to eliminate the prefactor of $L'(\alpha)$
and in the second step to obtain the expression for $C_{\upsilon}$.
Consequently, we have the FS
\begin{align}\label{SMDefXYFShv9}
\chi^{(h)}(h,\gamma)
&= \frac{\gamma^2}{4(1-\gamma^2)^2}\sum_{\upsilon=\pm}
\frac{\mathcal{F}_{\upsilon}^(h)}{(\lambda_{\upsilon}-\lambda_{\bar\upsilon})^2}
=\frac{1}{16(h^2+\gamma^2-1)}\sum_{\upsilon=\pm}\mathcal{F}_{\upsilon}^{(h)}.
\end{align}

\newpage
\subsection{Calculation for the anisotropy transition}
The formula for the FS of the anisotropy transition is
\begin{align}\label{SMDefXYFSrv1}
\chi^{(\gamma)}(h,\gamma)
&= \frac{1}{4(1-\gamma^2)^2}\sum_{n=1}^{N}\frac{(1-c_n^2)(c_n-h)^2}{(c_n-\lambda_{+})^2(c_n-\lambda_{-})^2}
\end{align}
where $c_n = \cos\frac{(2n-1)\pi}{\mathcal{N}}$~($\mathcal{N}=2N$) and
$\lambda_{\upsilon}$'s~($\upsilon=\pm$) are defined in Eq.~\eqref{SMlambdaNu}.
To eliminate the quadratic terms, we shall utilize the partial fraction expansion of the expression $\frac{(1-c_n^2)(c_n-h)^2}{(c_n-\lambda_{+})^2(c_n-\lambda_{-})^2}$.
It is known that for any $a$, $b$, and $c$, we have
\begin{equation*}
\frac{(1-x^2)(c+x)^2}{(x-a)^2(x-b)^2} = \mathcal{F}(a,b;c) + \mathcal{F}(b,a;c) - 1
\end{equation*}
where
\begin{equation*}
\mathcal{F}(a,b;c) = \frac{(1-a^2)(a+c)^2}{(a-b)^2(x-a)^2}
- \frac{2(a+c)\big[a(a-b)^2-(ab-1)(b+c)\big]}{(a-b)^3(x-a)}.
\end{equation*}
Therefore, Eq.~\eqref{SMDefXYFSrv1} can be recast into a symmetric form by $\mathcal{F}$.

We now simplify the summation
$\gamma^2\mathcal{F}_{\upsilon}^{(\gamma)} = (\lambda_{\upsilon}-\lambda_{\bar\upsilon})^2
\sum_{n=1}^{N}\mathcal{F}(\lambda_{\upsilon},\lambda_{\bar\upsilon}; -h)$,
which reads,
\begin{align*}
\gamma^2\mathcal{F}_{\upsilon}^{(\gamma)}
&= \sum\frac{(1-\lambda_{\upsilon}^2)(\lambda_{\upsilon}-h)^2}{(c_n-\lambda_{\upsilon})^2} - \sum\frac{2(\lambda_{\upsilon}-h)
\left[\lambda_{\upsilon}(\lambda_{\upsilon}-\lambda_{\bar\upsilon})^2-(\lambda_{\upsilon}\lambda_{\bar\upsilon}-1)(\lambda_{\bar\upsilon}-h)\right]}
{(\lambda_{\upsilon}-\lambda_{\bar\upsilon})(c_n-\lambda_{\upsilon})} \nonumber\\
&= (\lambda_{\upsilon}^2-1)(\lambda_{\upsilon}-h)^2L'(-\lambda_{\upsilon})
- \frac{2(\lambda_{\upsilon}-h)
\left[\lambda_{\upsilon}(\lambda_{\upsilon}-\lambda_{\bar\upsilon})^2-(\lambda_{\upsilon}\lambda_{\bar\upsilon}-1)(\lambda_{\bar\upsilon}-h)\right]}
{\lambda_{\upsilon}-\lambda_{\bar\upsilon}}L(-\lambda_{\upsilon}) \nonumber\\
&= (\lambda_{\upsilon}^2-1)(\lambda_{\upsilon}-h)^2L'(\lambda_{\upsilon})
+ \frac{2(\lambda_{\upsilon}-h)
\left[\lambda_{\upsilon}(\lambda_{\upsilon}-\lambda_{\bar\upsilon})^2-(\lambda_{\upsilon}\lambda_{\bar\upsilon}-1)(\lambda_{\bar\upsilon}-h)\right]}
{\lambda_{\upsilon}-\lambda_{\bar\upsilon}}L(\lambda_{\upsilon})
\end{align*}
where in the last step we have used the odevity of function $L(\alpha)$ and its derivative.
By virtue of the two Lemmas listed in the APPENDIX of the main text, we have
\begin{align*}
\mathcal{F}_{\upsilon}^{(\gamma)}
&= \frac{(\lambda_{\upsilon}^2-1)(\lambda_{\upsilon}-h)^2}{\gamma^2}
\frac{4g_{\upsilon}^2}{(g_{\upsilon}^2-1)^2}
    \left[\frac{\mathcal{N}^2g_{\upsilon}^{\mathcal{N}}}{(g_{\upsilon}^{\mathcal{N}}+1)^2}
    - \frac{\mathcal{N}}{2}
    \frac{(g_{\upsilon}^2+1)(g_{\upsilon}^{\mathcal{N}}-1)}{(g_{\upsilon}^2-1)(g_{\upsilon}^{\mathcal{N}}+1)}\right] \\
&\quad +\frac{2(\lambda_{\upsilon}-h)\left[\lambda_{\upsilon}(\lambda_{\upsilon}-\lambda_{\bar\upsilon})^2
    -(\lambda_{\upsilon}\lambda_{\bar\upsilon}-1)(\lambda_{\bar\upsilon}-h)\right]}
    {\gamma^2(\lambda_{\upsilon}-\lambda_{\bar\upsilon})}
    \frac{\mathcal{N}}{2\sqrt{\lambda_{\upsilon}^2-1}} \frac{g_{\upsilon}^{\mathcal{N}}-1}{g_{\upsilon}^{\mathcal{N}}+1} \\
&= \frac{(\lambda_{\upsilon}-h)^2}{\gamma^2}
\Bigg\{ \frac{\mathcal{N}^2g_{\upsilon}^{\mathcal{N}}}{(g_{\upsilon}^{\mathcal{N}}+1)^2} + \mathcal{N}
\left[\frac{\lambda_{\upsilon}(\lambda_{\upsilon}-\lambda_{\bar\upsilon})^2-(\lambda_{\upsilon}\lambda_{\bar\upsilon}-1)(\lambda_{\bar\upsilon}-h)}
    {(\lambda_{\upsilon}-h)(\lambda_{\upsilon}-\lambda_{\bar\upsilon})\sqrt{\lambda_{\upsilon}^2-1}} -\frac{g_{\upsilon}^2+1}{2(g_{\upsilon}^2-1)}\right]
\frac{g_{\upsilon}^{\mathcal{N}}-1}{g_{\upsilon}^{\mathcal{N}}+1} \Bigg\}  \\
&= \frac{(g_{\upsilon}^2-1)^2}{4g_{\upsilon}^2}
\Bigg\{ \frac{\mathcal{N}^2g_{\upsilon}^{\mathcal{N}}}{(g_{\upsilon}^{\mathcal{N}}+1)^2} + \frac{\mathcal{N}}{2\sqrt{\lambda_{\upsilon}^2-1}}
\left[\frac{2\lambda_{\upsilon}(\lambda_{\upsilon}-\lambda_{\bar\upsilon})^2-2(\lambda_{\upsilon}\lambda_{\bar\upsilon}-1)(\lambda_{\bar\upsilon}-h)}
    {(\lambda_{\upsilon}-h)(\lambda_{\upsilon}-\lambda_{\bar\upsilon})} - \lambda_{\upsilon}\right]
\frac{g_{\upsilon}^{\mathcal{N}}-1}{g_{\upsilon}^{\mathcal{N}}+1} \Bigg\}  \\
&= \frac{(g_{\upsilon}^2-1)^2}{4g_{\upsilon}^2}
\Bigg\{ \frac{\mathcal{N}^2g_{\upsilon}^{\mathcal{N}}}{(g_{\upsilon}^{\mathcal{N}}+1)^2} + \frac{\mathcal{N}}{2\sqrt{\lambda_{\upsilon}^2-1}}
\left[\frac{2(\lambda_{\upsilon}^2-1)(\lambda_{\upsilon}-\lambda_{\bar\upsilon})
-(\lambda_{\upsilon}^2+\lambda_{\upsilon}\lambda_{\bar\upsilon}-2)(\lambda_{\upsilon}-h)}
    {(\lambda_{\upsilon}-h)(\lambda_{\upsilon}-\lambda_{\bar\upsilon})} \right]
\frac{g_{\upsilon}^{\mathcal{N}}-1}{g_{\upsilon}^{\mathcal{N}}+1} \Bigg\}  \\
&= \frac{(g_{\upsilon}^2-1)^2}{4g_{\upsilon}^2}
\Bigg\{ \frac{\mathcal{N}^2g_{\upsilon}^{\mathcal{N}}}{(g_{\upsilon}^{\mathcal{N}}+1)^2} + \frac{\mathcal{N}}{2\sqrt{\lambda_{\upsilon}^2-1}}
\left[\frac{2(\lambda_{\upsilon}^2-1)}{\lambda_{\upsilon}-h} +
\frac{2-\lambda_{\upsilon}\lambda_{\bar\upsilon}-\lambda_{\upsilon}^2}{\lambda_{\upsilon}-\lambda_{\bar\upsilon}} \right]
\frac{g_{\upsilon}^{\mathcal{N}}-1}{g_{\upsilon}^{\mathcal{N}}+1} \Bigg\}  \\
&= \frac{(g_{\upsilon}^2-1)^2}{4g_{\upsilon}^2}
\Bigg[ \frac{\mathcal{N}^2g_{\upsilon}^{\mathcal{N}}}{(g_{\upsilon}^{\mathcal{N}}+1)^2} + \mathcal{N}
\left(\frac{p_{\upsilon}}{|\gamma|}+\frac{C_{\upsilon}}{2}\right)
\frac{g_{\upsilon}^{\mathcal{N}}-1}{g_{\upsilon}^{\mathcal{N}}+1} \Bigg]
\end{align*}
Here, \textit{Proposition 1} is used in the first step and is to eliminate part of the prefactor of $L'(\alpha)$
and in the second step to obtain the expression for $C_{\upsilon}$.
In the second step, \textit{Proposition 2} is also used.
Consequently, we have the FS
\begin{align}\label{SMDefXYFShv9}
\chi^{(\gamma)}(h,\gamma)
&= \frac{1}{4(1-\gamma^2)^2}\sum_{\upsilon=\pm}\frac{\gamma^2\mathcal{F}_{\upsilon}}{(\lambda_{\upsilon}-\lambda_{\bar\upsilon})^2}
-\frac{\mathcal{N}}{8(1-\gamma^2)^2}
=\frac{1}{16(h^2+\gamma^2-1)}\sum_{\upsilon=\pm}\mathcal{F}_{\upsilon}^{(\gamma)}-\frac{\mathcal{N}}{8(1-\gamma^2)^2}.
\end{align}

\newpage
\subsection{Calculation for the off-diagonal element of geometric tensor}
The formula for the off-diagonal element of geometric tensor is
\begin{align}\label{SMDefXYOffDiagv1}
\chi^{(h\gamma)}(h,\gamma)
&= \frac{\gamma}{4(1-\gamma^2)^2}\sum_{n=1}^{N}\frac{(1-c_n^2)(c_n-h)}{(c_n-\lambda_{+})^2(c_n-\lambda_{-})^2}
\end{align}
where $c_n = \cos\frac{(2n-1)\pi}{\mathcal{N}}$~($\mathcal{N}=2N$) and
$\lambda_{\upsilon}$'s~($\upsilon=\pm$) are defined in Eq.~\eqref{SMlambdaNu}.
To eliminate the quadratic terms, we shall utilize the partial fraction expansion of the expression $\frac{(1-c_n^2)(c_n-h)}{(c_n-\lambda_{+})^2(c_n-\lambda_{-})^2}$.
It is known that for any $a$, $b$, and $c$, we have
\begin{equation*}
\frac{(1-x^2)(c+x)}{(x-a)^2(x-b)^2} = \mathcal{F}(a,b;c) + \mathcal{F}(b,a;c)
\end{equation*}
where
\begin{equation*}
\mathcal{F}(a,b;c) = \frac{(1-a^2)(a+c)}{(a-b)^2(x-a)^2}
- \frac{a(a-b)^2-(ab-1)(a+b+2c)}{(a-b)^3(x-a)}.
\end{equation*}
Therefore, Eq.~\eqref{SMDefXYOffDiagv1} can be recast into a symmetric form by $\mathcal{F}$.

We now simplify the summation
$|\gamma|\mathcal{F}_{\upsilon}^{(h\gamma)} = (\lambda_{\upsilon}-\lambda_{\bar\upsilon})^2
\sum_{n=1}^{N}\mathcal{F}(\lambda_{\upsilon},\lambda_{\bar\upsilon}; -h)$,
which reads,
\begin{align*}
|\gamma|\mathcal{F}_{\upsilon}^{(h\gamma)}
&= \sum\frac{(1-\lambda_{\upsilon}^2)(\lambda_{\upsilon}-h)}{(c_n-\lambda_{\upsilon})^2}
- \sum\frac{\lambda_{\upsilon}(\lambda_{\upsilon}-\lambda_{\bar\upsilon})^2
-(\lambda_{\upsilon}\lambda_{\bar\upsilon}-1)(\lambda_{\upsilon}+\lambda_{\bar\upsilon}+2c)}
{(\lambda_{\upsilon}-\lambda_{\bar\upsilon})(c_n-\lambda_{\upsilon})} \nonumber\\
&= (\lambda_{\upsilon}^2-1)(\lambda_{\upsilon}-h)L'(-\lambda_{\upsilon})
- \frac{\lambda_{\upsilon}(\lambda_{\upsilon}-\lambda_{\bar\upsilon})^2
-(\lambda_{\upsilon}\lambda_{\bar\upsilon}-1)(\lambda_{\upsilon}+\lambda_{\bar\upsilon}+2c)}
{(\lambda_{\upsilon}-\lambda_{\bar\upsilon})}L(-\lambda_{\upsilon}) \nonumber\\
&= (\lambda_{\upsilon}^2-1)(\lambda_{\upsilon}-h)L'(\lambda_{\upsilon})
+ \frac{\lambda_{\upsilon}(\lambda_{\upsilon}-\lambda_{\bar\upsilon})^2
-(\lambda_{\upsilon}\lambda_{\bar\upsilon}-1)(\lambda_{\upsilon}+\lambda_{\bar\upsilon}+2c)}
{(\lambda_{\upsilon}-\lambda_{\bar\upsilon})}L(\lambda_{\upsilon})
\end{align*}
where in the last step we have used the odevity of function $L(\alpha)$ and its derivative.
By virtue of the two Lemmas listed in the APPENDIX of the main text, we have
\begin{align*}
\mathcal{F}_{\upsilon}^{(h\gamma)}
&= \frac{(\lambda_{\upsilon}^2-1)(\lambda_{\upsilon}-h)}{|\gamma|}
\frac{4g_{\upsilon}^2}{(g_{\upsilon}^2-1)^2}
    \left[\frac{\mathcal{N}^2g_{\upsilon}^{\mathcal{N}}}{(g_{\upsilon}^{\mathcal{N}}+1)^2}
    - \frac{\mathcal{N}}{2}
    \frac{(g_{\upsilon}^2+1)(g_{\upsilon}^{\mathcal{N}}-1)}{(g_{\upsilon}^2-1)(g_{\upsilon}^{\mathcal{N}}+1)}\right] \\
&\quad +\frac{\lambda_{\upsilon}(\lambda_{\upsilon}-\lambda_{\bar\upsilon})^2
    -(\lambda_{\upsilon}\lambda_{\bar\upsilon}-1)(\lambda_{\upsilon}+\lambda_{\bar\upsilon}+2c)}
    {|\gamma|(\lambda_{\upsilon}-\lambda_{\bar\upsilon})}
    \frac{\mathcal{N}}{2\sqrt{\lambda_{\upsilon}^2-1}} \frac{g_{\upsilon}^{\mathcal{N}}-1}{g_{\upsilon}^{\mathcal{N}}+1} \\
&= \frac{\lambda_{\upsilon}-h}{|\gamma|}
\Bigg\{ \frac{\mathcal{N}^2g_{\upsilon}^{\mathcal{N}}}{(g_{\upsilon}^{\mathcal{N}}+1)^2} + \mathcal{N}
\left[\frac{\lambda_{\upsilon}(\lambda_{\upsilon}-\lambda_{\bar\upsilon})^2-(\lambda_{\upsilon}\lambda_{\bar\upsilon}-1)(\lambda_{\upsilon}+\lambda_{\bar\upsilon}+2c)}
    {2(\lambda_{\upsilon}-h)(\lambda_{\upsilon}-\lambda_{\bar\upsilon})\sqrt{\lambda_{\upsilon}^2-1}} -\frac{g_{\upsilon}^2+1}{2(g_{\upsilon}^2-1)}\right]
\frac{g_{\upsilon}^{\mathcal{N}}-1}{g_{\upsilon}^{\mathcal{N}}+1} \Bigg\}  \\
&= \frac{g_{\upsilon}^2-1}{2p_{\upsilon}g_{\upsilon}}
\Bigg\{ \frac{\mathcal{N}^2g_{\upsilon}^{\mathcal{N}}}{(g_{\upsilon}^{\mathcal{N}}+1)^2} + \frac{\mathcal{N}}{2\sqrt{\lambda_{\upsilon}^2-1}}
\left[\frac{\lambda_{\upsilon}(\lambda_{\upsilon}-\lambda_{\bar\upsilon})^2-(\lambda_{\upsilon}\lambda_{\bar\upsilon}-1)(\lambda_{\upsilon}+\lambda_{\bar\upsilon}+2c)}
    {(\lambda_{\upsilon}-h)(\lambda_{\upsilon}-\lambda_{\bar\upsilon})} - \lambda_{\upsilon}\right]
\frac{g_{\upsilon}^{\mathcal{N}}-1}{g_{\upsilon}^{\mathcal{N}}+1} \Bigg\}  \\
&= \frac{g_{\upsilon}^2-1}{2p_{\upsilon}g_{\upsilon}}
\Bigg\{ \frac{\mathcal{N}^2g_{\upsilon}^{\mathcal{N}}}{(g_{\upsilon}^{\mathcal{N}}+1)^2} + \frac{\mathcal{N}}{2\sqrt{\lambda_{\upsilon}^2-1}}
\left[\frac{(\lambda_{\upsilon}^2-1)(\lambda_{\upsilon}-\lambda_{\bar\upsilon})
-(\lambda_{\upsilon}^2+\lambda_{\upsilon}\lambda_{\bar\upsilon}-2)(\lambda_{\upsilon}-h)}
    {(\lambda_{\upsilon}-h)(\lambda_{\upsilon}-\lambda_{\bar\upsilon})} \right]
\frac{g_{\upsilon}^{\mathcal{N}}-1}{g_{\upsilon}^{\mathcal{N}}+1} \Bigg\}  \\
&= \frac{g_{\upsilon}^2-1}{2p_{\upsilon}g_{\upsilon}}
\Bigg\{ \frac{\mathcal{N}^2g_{\upsilon}^{\mathcal{N}}}{(g_{\upsilon}^{\mathcal{N}}+1)^2} + \frac{\mathcal{N}}{2\sqrt{\lambda_{\upsilon}^2-1}}
\left[\frac{\lambda_{\upsilon}^2-1}{\lambda_{\upsilon}-h} +
\frac{2-\lambda_{\upsilon}\lambda_{\bar\upsilon}-\lambda_{\upsilon}^2}{\lambda_{\upsilon}-\lambda_{\bar\upsilon}} \right]
\frac{g_{\upsilon}^{\mathcal{N}}-1}{g_{\upsilon}^{\mathcal{N}}+1} \Bigg\}  \\
&= \frac{g_{\upsilon}^2-1}{2p_{\upsilon}g_{\upsilon}}
\Bigg[ \frac{\mathcal{N}^2g_{\upsilon}^{\mathcal{N}}}{(g_{\upsilon}^{\mathcal{N}}+1)^2} + \frac{\mathcal{N}}{2}
\left(\frac{p_{\upsilon}}{|\gamma|}+C_{\upsilon}\right)
\frac{g_{\upsilon}^{\mathcal{N}}-1}{g_{\upsilon}^{\mathcal{N}}+1} \Bigg]
\end{align*}
Here, \textit{Proposition 1} is used in the first step and is to eliminate part of the prefactor of $L'(\alpha)$
and in the second step to obtain the expression for $C_{\upsilon}$.
In the second step, \textit{Proposition 2} is also used.
Consequently, we have the FS
\begin{align}\label{SMDefXYOffDiagv9}
\chi^{(h\gamma)}(h,\gamma)
&= \frac{\gamma}{4(1-\gamma^2)^2}\sum_{\upsilon=\pm}\frac{|\gamma|\mathcal{F}_{\upsilon}^{(h\gamma)}}{(\lambda_{\upsilon}-\lambda_{\bar\upsilon})^2}
=\frac{\textrm{sgn}(\gamma)}{16(h^2+\gamma^2-1)}\sum_{\upsilon=\pm}\mathcal{F}_{\upsilon}^{(h\gamma)}.
\end{align}

\end{document}